\documentclass[prl,aps,reprint]{revtex4-1}
\usepackage{graphicx}
\usepackage{amsmath,amssymb}
\usepackage[caption=false]{subfig}
\usepackage[breaklinks=true,colorlinks=true,bookmarks=false,urlcolor=blue,
citecolor=blue,linkcolor=blue]{hyperref}

\begin{document}

\title{Breakup of liquid jets: Thermodynamic perspectives}

\author{Fei Wang$^{1,5}$}
\email{fei.wang@kit.edu}
\author{Oleg Tschukin$^{1}$}
\author{Thomas Leisner$^{3,4}$}
\author{Haodong Zhang$^{1,5}$}
\author{Britta Nestler$^{1,2,5}$}

\affiliation{$^{1}$Institute of Applied Materials, 
 Karlsruhe Institute of Technology (KIT),
                     Stra\ss e am Forum 7, 76131 Karlsruhe, Germany}
                     
\affiliation{$^{2}$Institute of Digital Materials Science, 
Karlsruhe University of Applied Sciences,
Moltkestra{\ss}e 30, 76133 Karlsruhe, Germany}

\affiliation{$^{3}$Institute
 of Environmental Physics, 
 University of Heidelberg, Im Neuenheimer Feld 229, 69120 Heidelberg, Germany}

\affiliation{$^{4}$Institute
 of Meteorology and Climate Research, Karlsruhe Institute of Technology (KIT), 
 Hermann-von-Helmholtz-Platz 1,
76344 Eggenstein-Leopoldshafen, Germany}

\affiliation{$^{5}$Institute of Nanotechnology, Karlsruhe Institute of Technology (KIT), 
Hermann-von-Helmholtz-Platz 1, 76344 Eggenstein-Leopoldshafen, Germany}

  \date{\today}
\begin{abstract}
Breakup of a liquid jet into a chain of droplets is common in 
 nature and industry.
 Previous researchers developed profound mathematic and fluid dynamic
 models to address 
 this breakup phenomenon starting from tiny perturbations.
 However, the morphological evolution of the jets at large amplitude perturbations
 is still an  open question.
 Here, we report a concise thermodynamic model 
 based on the surface area minimization principle.
 Our results demonstrate
 a reversible breakup transition from a continuous jet via droplets towards 
 a uniform-radius cylinder,  which has not been found previously,
 but is observed in our simulations. 
 This new observation is attributed to a 
 geometric constraint, which was often overlooked in former studies.
 We anticipate our model to be a shortcut
 to tackle many similar highly nonlinear morphological evolutions 
 without solving abstruse fluid dynamic equations for inviscid fluids.
\end{abstract}
\maketitle

\begin{figure}[h!]
	\centering
	\includegraphics[width=1\linewidth]{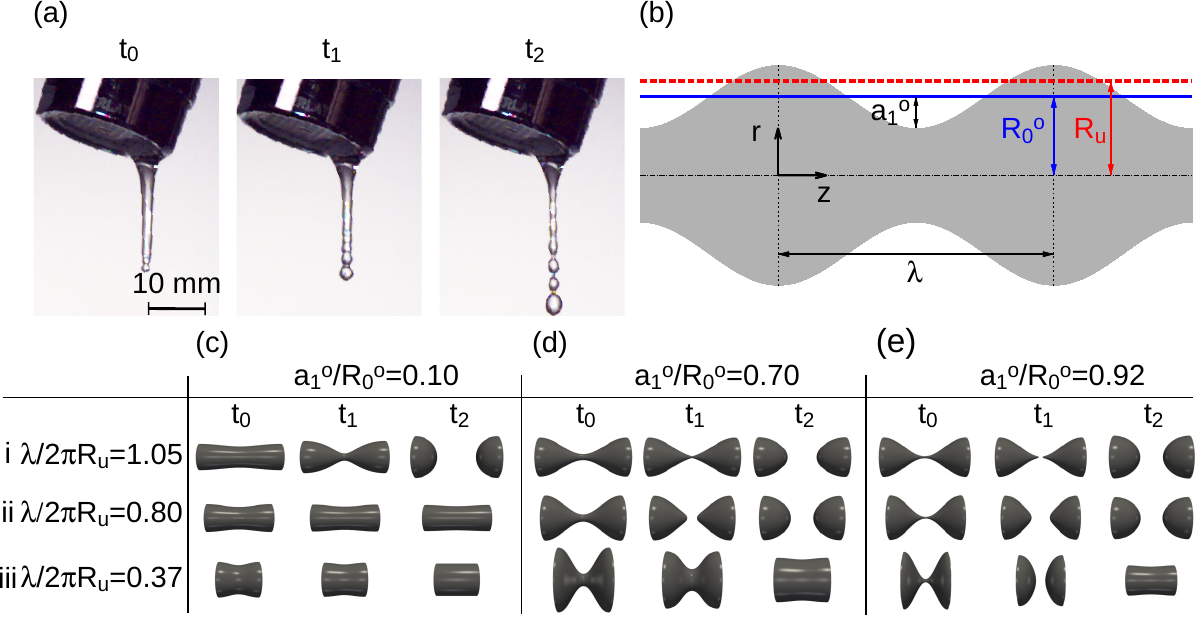}	
	\caption{\textbf{Morphological evolution of liquid jets.}
        (a) Formation of droplets when water trickles down from a water-tap.
	(b) 2D projection of a perturbed jet whose surface follows
		$r_o=R_0^o+a_1^o\cos(2\pi z/\lambda)$, where
		$a_1^o$ and $\lambda$ 
		are the amplitude and the wavelength of the initial harmonic perturbation, 
		respectively.
		The volume of the jet 
		is $\int_0^\lambda \pi r_o^{2} dz=\pi [(R_0^{o})^2+\frac{1}{2}(a_1^{o})^2]\lambda$
		and we define $R_u^2=(R_0^{o})^2
		+\frac{1}{2}(a_1^{o})^2$ as the mean radius 
		of the jet.
		(c), (d), (e), Morphological evolution 
		of jets with different amplitudes and wavelengths  
		via phase-field simulations.
		All the results in (c) (tiny perturbations) 
		coincide with Rayleigh's theory~\cite{Rayleigh1879}.
		Conversely, in (d)-ii, 
		the breakup wavelength
		deviates from Rayleigh's prediction and this deviation
		also appears in (e)-ii.
		In (e)-iii, we observe an unusual breakup: 
		a perturbed jet $\rightarrow$ dispersed droplets 
		$\rightarrow$ continuous cylinder. 
	}
	\label{fig:1}
\end{figure}

As shown in Fig.~\ref{fig:1}(a),  when we turn on a water-tap, 
a water jet trickles down and 
eventually breaks apart into a chain of 
droplets, which is also observed in the falling rain.
This breakup phenomenon 
has drawn broad interests both in 
fundamental 
researches~\cite{kaufman2012structured,passian2012materials,royer2009high,
loscertales2002micro,utada2005monodisperse,moseler2000formation,haefner2015influence,cardoso2006rayleigh,prado2011experimental,wang2020acta,wang2016scritpa}  
and practical applications~\cite{wijshoff2010dynamics,bhat2010formation,
xue2016engineering,garnett2012self,day2015plateau}.
In the 19th century, Joseph Plateau and 
Lord Rayleigh~\cite{Rayleigh1878,Rayleigh1879} proposed 
a quintessential theory 
for droplet-formation considering tiny perturbations.
However, the prevalence of large fluctuations
in nature impedes the adaptability of this tenet.
 Hence, finding a satisfactory answer to the classic Plateau-Rayleigh 
 question holds great promise not only for explicating
 a vast amount of unexplainable experiments~\cite{boukharov2008dynamics}
but also to guide and improve advanced technical 
 applications. 

Here, we report a concise mathematical-physical model to address 
the formation of droplets via the breakup of a liquid jet from 
a thermodynamic point of view. 
Traditionally,  the Plateau-Rayleigh question was 
tackled by solving intricate fluid dynamic 
equations~\cite{yuen1968non,nayfeh1970nonlinear}. 
Ipso facto, the occurrence of several irregular breakup phenomena 
via the phase-field model (\textcolor{blue}{Fig.~\ref{fig:1}(d)-ii,
1(e)-ii and 1(e)-iii}) sheds light on 
the ambiguity of those conventional treatments.
Presently, rather than pondering Navier-Stokes equations,  
we tackle the classic Plateau-Rayleigh question 
by considering the temporal dissipation of the Gibbs free energy of the system.

\begin{figure*}[hbt!]
	\centering
		\includegraphics[width=0.85\linewidth]{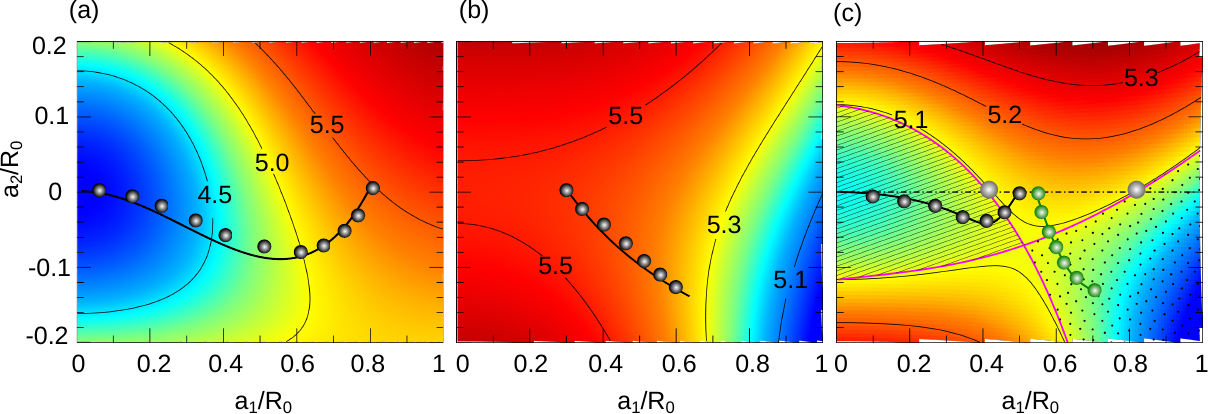}
	\caption{\textbf{Surface area landscape.}
		(a), (b), (c), 
		Surface area of jets as 
		function of all the possible values of the
		first two Fourier coefficients ($a_1/R_0$ and $a_2/R_0$)
		with normalized wavelengths 
		$\lambda/(2\pi R_u)=0.37$ 
		(Fig.~\ref{fig:1}iii), $1.05$ (Fig.~\ref{fig:1}i), 
		$0.80$ (Fig.~\ref{fig:1}ii), respectively.	
		The black/green circles denote
		 the evolution routes of $a_1/R_0$ and $a_2/R_0$ 
		 from the phase-field simulations.
		 The black/green dashed curves 
		 represent the  evolution paths 
		 from the gradient descent method.		
		The gray circles in (c) embrace 
		a barrier interval along the horizontal dot-dashed line $a_2/R_0=0$.
		The hatched and dotted regions in (c)  are 
		partitioned by the isolines (magenta lines) of
		the saddle point of the surface area landscape.
	}
	\label{fig:2}
\end{figure*}
Our consideration is based on the thermodynamic principle of surface area minimizing 
with time irrespective of kinetics.
Most importantly, one crucial fact is always overlooked that 
the initially cosinusoidal jet is getting disordered gradually with time, 
in lieu of remaining harmonic~\cite{carter1987effect,nichols1965surface}. 
Hence, we employ the following  Fourier series~\cite{wang2020acta}
\begin{equation*}
 r(t,z)=R_0(t)+\sum_{n=1}^K a_n(t)\cos nkz,
 \label{eq:1}
\end{equation*}
to depict the instantaneously changing jet-surface more precisely
prior to the breakup with the initial condition: 
$a_1(t=0)=a_1^o$,  $a_n(t=0)=0,~n\geq2$ and $R_0(t=0)=R_0^o$. Here, 
 $k=2\pi/\lambda$ is the wavenumber,
 $a_n$ are the Fourier coefficients, 
 and $K$ is the dimension of the Fourier series.
 It is noteworthy that the Fourier coefficients $a_n(t)$ 
 are constrained  by the volume conservation
as $V(t)= \pi R_u^2\lambda$, which yields
$R_u^2=R_0^2(t)+\frac{1}{2}\sum_{n=1}^Ka_n^2(t),\forall t$.
The surface area of the jet is calculated by the following integral 
(see details in Refs.~\cite{wang2020acta,wang2016scritpa})
\begin{align*}
 S(a_1(t),a_2(t),\cdots)=&2\pi \int_0^\lambda r(t,z) \sqrt{1+[\partial_z r(t,z)]^2} dz.
 \end{align*}

\begin{figure*}[hbt!]
	\centering
		\includegraphics[width=0.8\linewidth]{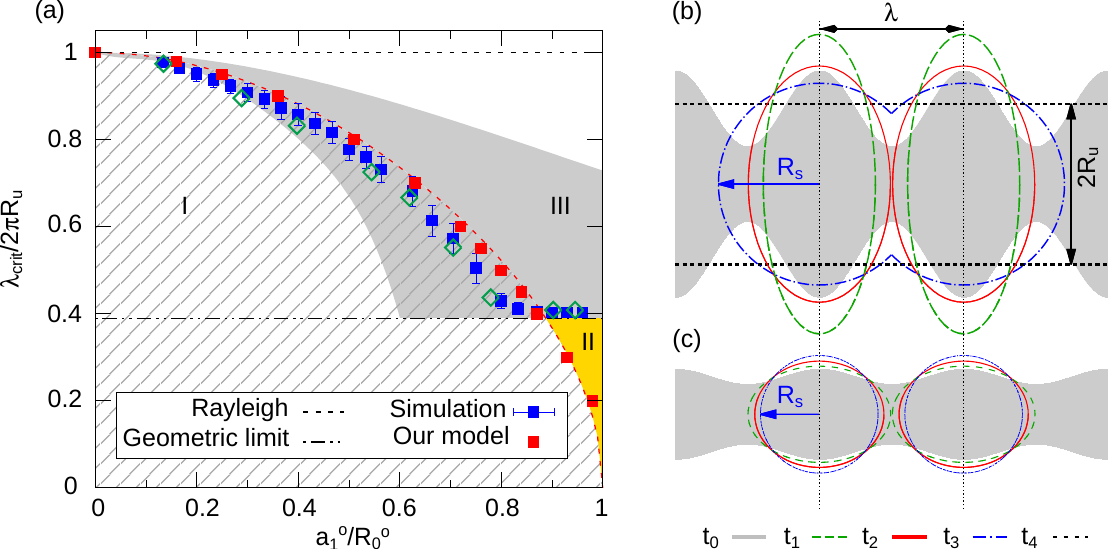}
	\caption{\textbf{Stability diagram.}
		(a) The normalized critical breakup wavelength
		$\lambda_{\text{crit}}/(2\pi R_\text{u})$
		as a function of the scaled initial amplitude $a_1^o/R_0^o$.
		The red and blue squares depict
		the results from the gradient descent method and 
		the Allen-Cahn model, respectively. 
			The red dashed line is the fitting curve for the red squares.
	The green diamonds demonstrate the simulation results from
	a fluid dynamics model (see supplemental document), which are in 
	good agreement with our model.
		 The dot-dashed line denotes 
the geometric criterion.
The gray shaded region
 illustrates all the barrier intervals shown in Fig.~\ref{fig:2}(c) for different wavelengths. 
		(b)  
		A reversible separation in II
		from a continuous jet via separated
		ellipsoid-shaped droplets towards a uniform-radius cylinder. 
		(c) Regular breakup in III, where $R_s$ is the radius of the 
		resulting spheroids.}
	\label{fig:3}
\end{figure*}

For a given wavelength $\lambda$,  
we visualize the surface area landscape $S(a_1,a_2)$
for all possible values of the two Fourier coefficients $a_1$ and $a_2$. 
Three typical surface area landscapes $S$ as a function of 
$a_1$ and $a_2$ 
are illustrated in Fig.~\ref{fig:2}(a), (b) and (c)
for short, long, and medium wavelengths, respectively.
For the short (long) wavelength,
the global minimum  of the
 surface area is
 at $a_1=0$ (nearby the maximal value of $a_1$),
 so that the end-state is a uniform-radius cylinder (droplets). 
Interestingly, for the intermediate wavelength (Fig.~\ref{fig:2}(c)), 
two local minima locate inside the hatched and
the dotted regions, which correspond to a final state of a 
uniform-radius cylinder and  a droplet-structure, respectively.
The domain outside these two regions, which is called as 
barrier zone, highlights initial configurations
whose surface area is greater than the ones in the hatched and dotted 
regions and can evolve either into a uniform-radius
cylinder or into droplets. 
As aforementioned, we consider 
the evolution of a jet with a cosinusoidal perturbation with $a_2^o=0$ at the beginning.
This initial setup corresponds to the horizontal dot-dashed line in Fig.~\ref{fig:2}(c)
and overlaps the barrier zone between the two gray circles 
embracing a barrier interval.
The critical configuration for the droplet-formation
buries inside this barrier interval, which
is also obtained for many other medium wavelengths.
Those barrier intervals versus wavelengths
is represented by the shaded region in 
Fig.~\ref{fig:3}(a). 
Left below and right above this shaded region, 
the jet has a surface area landscape as in Fig.~\ref{fig:2}(a) and 2(b), respectively.

Next, we adopt the phase-field simulations~\cite{wang2012,wang2019acta,Laxmipathy2021,Laxmipathy2021cms}
and the gradient descent method (GDM)~\cite{wu2019pre} to
scrutinize the evolution kinetics and 
the critical breakup configuration. 
In the phase-field model, 
a phase order parameter $\varphi$ is introduced to characterize the phase state. 
For instance, $\phi=1$ denotes
the jet,  $\phi=0$ represents the surrounding, and 
$\phi$ continuously varies from 1 to 0 in the diffuse interface from the jet to the surrounding.
The evolution of the phase order parameter $\partial_t\varphi$
is such as to reduce the Ginzburg-Landau energy functional, $\mathcal S(\varphi)$, 
following GDM as
\begin{equation*}
 \partial_t\varphi=-\delta \mathcal S/\delta \varphi-\mathcal L.
\end{equation*}
Here, $\mathcal L$ is a Lagrange multiplier to ensure the volume conservation.
The Ginzburg-Landau energy functional is expressed as~\cite{cahn1958free}
$\mathcal S(\varphi)=\int_\Omega [\kappa (\nabla\varphi)^2+\omega(\varphi)] d\Omega$,
where $\kappa$ is the gradient energy coefficient and 
$\omega(\varphi)$ denotes
an obstacle potential energy (see supplemental document for details). 

Since surface tension is a joint effect of adhesion and cohesion,
which both are conservative forces, the evolution direction of the Fourier coefficients
follows the gradient of the surface area $-(\partial_{a_n}S)\in \mathbb R^K$,
namely,
\begin{equation*}
 \partial_ta_n=-\partial_{a_n}S.
\end{equation*}
The time evolution is subjected to the aforementioned initial condition.
In Fig.~\ref{fig:2}(a) and 2(b), the black solid lines depict
the kinetic routes along the steepest gradient for two 
initial setups $a_1^o/R_0^o=0.8$ and $0.3$, respectively.
The GDM paths are well consistent with the simulation results (black circles),
which are achieved by applying the
Fourier transformation to the jet-surface
to obtain the  Fourier coefficients at different time steps of the simulation.
In Fig.~\ref{fig:2}(c), the black and green lines (GDM)/circles (simulations)
 illustrate
the evolution paths of two exemplary setups inside the barrier interval, 
$a_1^o/R_0^o=0.48$ and $0.55$,
which transform into a uniform-radius cylinder
and a droplet-structure, respectively.
By using binary search algorithm 
for initial setups inside the barrier interval, the 
critical breakup configurations are identified by GDM and simulations,
as depicted by the red and blue squares, respectively, in Fig.~\ref{fig:3}(a).

As shown in Fig.~\ref{fig:3}(a), our simulation 
results coincide quite well with GDM
when $a_1^o/R_0^o\lesssim0.88$.
While $a_1^o/R_0^o>0.88$, 
the simulation results become a horizontal line
which surprisingly deviates from  GDM.
A heedful scrutiny on those unusual simulations 
reveals that the jet indeed firstly breaks apart into several 
oval-shaped droplets (see Fig.~\ref{fig:1}(e)-iii, t$_1$)
in accordance
with  GDM. 
However, afterwards, the spheroidization of the oval droplets
rebuilds contact between neighbors
and finally leads to  a uniform-radius cylinder.
This process is sketched in Fig.~\ref{fig:3}(b)
and decided by a geometric limit
where the
distance between the centroids 
of the resulting spheroids $2R_s$
is equal to the wavelength of the perturbation.
With the volume conservation condition $\frac{4}{3}\pi R_s^3=\pi R_u^2 \lambda$,
the geometric criterion is derived: 
$\lambda_{\text{crit}}=\sqrt{6} R_u$,
which is shown by
the horizontal dot-dashed line in Fig.~\ref{fig:3}(a) and provides 
a reasonable interpretation for those abnormal simulative morphological evolutions 
deviating from GDM.

As a result of the geometric constraint,
the stability diagram in Fig.~\ref{fig:3}(a) is divided into three regimes:
I (hatch line), II (orange) and III (excluding I and II).
In I, the perturbed jet 
directly evolves into a uniform-radius cylinder.
In II, the jet firstly transforms into 
separated prolate spheroids, elongated in the radial dimension, as shown by the 
green line in Fig.~\ref{fig:3}(b).
Afterwards, spheroidization 
rebuilds a chain of connected droplets (red and blue lines),
which eventually evolves into a uniform-radius cylinder (black dashed line).
In III, the jet also decomposes into ellipsoid-shaped droplets,
which are, however, oblate this time. 
As schematically shown  in Fig.~\ref{fig:3}(c),
the decrease in the surface area of the oblate spheroids
results in an augmentation of the gap spacing between adjacent droplets.
The demarcation between II and III
is defined by the locus that 
neither prolate nor oblate, but spherical droplets are precisely tangent to their neighbors.
This critical configuration is actually the geometric limit mentioned above.

In conclusion, we have proposed a thermodynamic
concept to address the classic Plateau-Rayleigh question.  
The surface area landscape method
combining with the gradient descent approach
as well as the geometric constraint
enables us to reveal a transient reversible breakup 
phenomenon. 
Furthermore, precise knowledge of non-linear
stability analysis for large perturbations
 is still lacking at present, 
 which makes it difficult to gain
 efficient predictions for highly non-linear morphological evolutions. 
We anticipate that the surface area landscape method proposed in the present work
 is an alternative for nonlinear stability analysis
 to see through more physical phenomena.

 
\textbf{Acknowledgments:}
The authors thank for support through the coordinated program 
Virtual Materials Design
(VirtMat) within the Helmholtz association and through the Gottfried-Wilhelm Leibniz
programme NE 822/31. F.W. is grateful to the discussion with W. C. Carter (MIT).
F.W. designed the phase-field simulations.
F.W. and  O.T. contributed to the concept 
of the surface area landscape.  F.W. and  H.Z. discussed the gradient descent method
and wrote the first draft of the manuscript. 
T.L. contributed to the picture illustrating the breakup of a liquid jet under water tap.
T.L. and B.N. interpreted the thermodynamic concept 
and revised the manuscript. 
All authors contributed to the manuscript preparation.

\end{document}